\documentclass{ws-p8-50x6-00}
\input epsf.sty

\newcommand{\ba}{\begin{eqnarray}}
\newcommand{\Tr}{{\rm Tr\,}}
\newcommand{\im}{\mathop{\rm Im}}
\newcommand{\re}{\mathop{\rm Re}}
\newcommand{\ea}{\end{eqnarray}}
\newcommand{\rmi}[1]{{\mbox{\scriptsize #1}}}
\def\lsi{\raise0.3ex\hbox{$<$\kern-0.75em\raise-1.1ex\hbox{$\sim$}}}
\def\gsi{\raise0.3ex\hbox{$>$\kern-0.75em\raise-1.1ex\hbox{$\sim$}}}
\newcommand{\lsim}{\mathop{\lsi}}
\newcommand{\gsim}{\mathop{\gsi}}

\begin{document}

\title{Electroweak phase transition\\ beyond the Standard Model}

\author{M. Laine}

\address{Theory Division, CERN, CH-1211 Geneva 23, Switzerland\\
E-mail: mikko.laine@cern.ch}

\maketitle

\abstracts{Standard theories of electroweak interactions 
are based on the concept of a gauge symmetry broken by the Higgs
mechanism. If they are placed in an environment with a sufficiently 
high temperature, the symmetry gets restored. It turns out that 
the characteristics of the symmetry restoring phase transition, 
such as its order, are important for cosmological applications, 
such as baryon asymmetry generation. We first briefly review how, 
by a combination of analytic and numerical methods, the properties of 
the phase transition can be systematically resolved for 
any given type of a (weakly interacting) Higgs sector. We then summarise 
the numerical results available for the Standard Model, and present 
a generic model independent statement as to how the Higgs sector should 
differ from the Standard Model for the properties of the transition 
to be very different. As an explicit example, we discuss the 
possibilities available for a strong transition
in the experimentally allowed parameter region of 
the Minimal Supersymmetric Standard Model.}

\vspace*{-9.5cm}
\begin{flushright} hep-ph/0010275
\end{flushright}
\vspace*{9.0cm}

\section{Introduction}

Even though a Higgs particle remains to be discovered, the 
success of precision computations gives strong support to 
the idea of an SU(2)$\times$U(1) electroweak theory with one or more 
Higgs doublets, some of which have, in any practical gauge, a non-vanishing 
expectation value at zero temperature. However, when the theory is placed 
in an environment with a sufficiently high temperature, the symmetry 
generically gets restored.~\cite{m0l:early} Thus we may imagine that 
there was an electroweak phase transition in the Early Universe.

An outstanding physics motivation for these considerations 
was perceived about fifteen years ago.~\cite{m0l:krs} Indeed,
an electroweak phase transition might in principle leave as a remnant 
the matter--antimatter (baryon) asymmetry presently observed. An essential
ingredient in this statement is that there are anomalous
baryon number violating interactions which are fast enough to 
be in thermal equilibrium at temperatures above the transition.~\cite{m0l:krs} 
If the transition is sufficiently strongly of the first order, i.e. 
discontinuous, baryon number violating interactions are however
switched off below the transition temperature, 
in the broken phase. Together with the CP violation known 
to exist in nature, as well as the non-equilibrium dynamics always
present in first order transitions, these ingredients might suffice
to produce an asymmetry with the correct 
order of magnitude (for a review, see ref.~\cite{m0l:rs}).

In the present talk, we will not enter any details of the actual 
baryogenesis computations, but simply consider the existence and 
strength of a first order phase transition. As a reference value, 
we however recall that the constraint of sufficient baryon violation
switch-off after the transition can be related to 
the Higgs expectation value over temperature $v/T$ in, say, the 
Landau gauge, or gauge invariantly to an appropriately regularized
value of $\langle \phi^\dagger \phi\rangle /T^2\approx v^2/(2T^2)$, 
where $\phi$ 
is the Higgs doublet. In the case of several SU(2)
Higgs doublets, the observable is correspondingly 
$\sum_i \langle\phi^\dagger_i \phi_i \rangle/T^2$.
The numerical requirement, based on 1-loop
saddle point computations (for a summary of their status and 
accuracy, see~\cite{m0l:nonpert}) as well as on a non-perturbative
evaluation,~\cite{m0l:moore_broken} is that the 
discontinuity in $v/T$ across the phase 
transition should exceed unity, $\Delta v/T \gsim 1.0$.

\section{General formulation of the problem}

Let us start by stating more generally the types of quantities 
that we may want to determine. The most basic characteristic
of a finite temperature system is its partition 
function $Z(T,V)$, or the free energy density $f(T)$, where
\be
Z(T,V) = e^{-\beta {V f(T)}} = \Tr e^{-\beta {H}}. 
\ee
Here $H$ is the Hamiltonian, $\beta = T^{-1}$, 
and $V\to\infty$ is the volume.
The points $T=T_c$ where $f(T)$ is non-analytic are
phase transition points, 
and if $\left. f'(T)\right|_{T\to T_c^\pm}$ is discontinuous, the 
transition is of the first order. This implies also that 
the energy density, $e(T) = -T^2 d[T^{-1} f(T)]/dT$, 
is discontinuous. 
In addition to $f(T),e(T)$, we also want to determine, 
as discussed above, the value of a (suitably regularized) 
$\langle \phi^\dagger\phi \rangle (T)$, which is very strongly 
correlated with the rate of baryon number violation
in the broken phase. 

In perturbation theory, one does not 
usually address directly $f(T)$ or 
$\langle\phi^\dagger\phi\rangle$, but 
rather the effective potential $V(v,T)$
for the length of the Higgs field, 
$|\phi| = v/\sqrt{2}$. After gauge fixing, 
$V(v,T)$ can be defined for instance
as the free energy density after the introduction of a constraint 
$\delta (v/\sqrt{2} - V^{-1} \int d^3x |\phi(x)| )$  in the path integral. 
In the thermodynamic limit, the value of $V(v,T)$ at the global 
minimum gives the actual free energy density, $f(T)$. 

We may now recall that
for all such ``static'' observables, it 
is straightforward to write down an ``imaginary time'' Euclidian 
path integral expression. Thus, in principle
we only need to compute
simple well-defined functional integrals, 
without any ambiguities related e.g.\ to Minkowski space.

\section{Why is it non-trivial?}

One may ask, why not simply trust $V(v,T)$ computed from
the functional integral using
perturbation theory? Indeed, such computations have been 
carried out up to resummed 2-loop level for the Standard 
Model (SM)~\cite{m0l:ae,m0l:fh,m0l:pert}, as well as for the 
MSSM~\cite{m0l:e}$^{-}$\cite{m0l:mlo3}. 
It turns out that at finite temperatures
perturbative computations are not a priori reliable, however. 
We will meet circumstances below where even the order of the 
phase transition is wrong.  

The reason for the 
breakdown is referred to as the infrared (IR)
problem of finite temperature field theory.~\cite{m0l:gpy}
It concerns ``light'' bosonic degrees of freedom, and can be understood
as arising from expansion parameters of the type
\be
g^2 n_b(m) = \frac{g^2}{e^{m/T}-1} \stackrel{m\ll T}{\sim}\frac{g^2T}{m},
\ee
where $n_b(m)$ is the Bose-Einstein distribution function, 
$g^2$ is a generic coupling constant, the SU(2) gauge coupling
in particular, and $m$ 
some perturbative mass scale appearing in the 
propagators. Adding numerical factors from known contributions
up to 3-loop order,
the largest expansion parameter in the system
can more concretely be estimated as~\cite{m0l:nonpert}
\be
\epsilon\approx
\frac{g^2 T}{\pi m}. \label{m0l:epsilon}
\ee
Thus, relatively ``light'' ($m\ll g^2 T$) degrees of 
freedom interacting with the Higgses are a problem,
and should be studied non-perturbatively. Such modes 
include, for instance, SU(2) gauge bosons near the symmetric phase.

In principle, a seemingly obvious way to study the system
non-pertur\-ba\-tively is to carry out four-dimensional
(4d) finite temperature lattice simulations. However,
in the present context this turns out to be quite 
demanding. This is, first of all, because the coupling is 
weak, so that there are multiple length
scales, $(2 \pi T)^{-1}, (gT)^{-1} , (g^2T)^{-1}$, 
difficult to fit simultaneously on a finite lattice. 
The second problem is that only the bosonic 
sector of the theory can be studied anyway, 
since chiral fermions cannot be efficiently handled in practice. 
Nevertheless, leaving out the fermions, large scale efforts
have been undertaken, complete with an 
extrapolation to the continuum limit 
(\cite{m0l:cfh,m0l:4dmssm} and references therein).
The method we shall follow here is, however, different. 

\section{Dimensional reduction and 3d effective field theories}

It turns out that all the problems mentioned above can be 
overcome in an economic and precise way
by the method of finite temperature
effective field theories (for reviews, 
see~\cite{m0l:erice}). 
The basic idea of such an approach
is to combine perturbation theory and simulations, in 
regions where they work best: one can integrate out massive modes 
perturbatively, 
which works well since the couplings 
are small, and then study the dynamics of the light modes 
non-perturbatively.
In the first step, the original 4d theory
reduces to a three-dimensional (3d) effective one. For studying the 
electroweak phase transition in a weakly coupled 
theory ($m_H\lsim 250$ GeV), 
this approach works with a practical accuracy at the percent 
level, both from the point of view of dimensional 
reduction,~\cite{m0l:cfh,m0l:generic,m0l:gr2} as well as numerical 
simulations.~\cite{m0l:nonpert,m0l:su2u1,m0l:owe,m0l:gis}

More concretely, the degrees of freedom integrated 
out~\cite{m0l:pert,m0l:generic,m0l:jp,m0l:mssmold,m0l:ll} 
are all the heavy beyond-the-SM particles, as well 
as the non-zero Matsubara modes of the SM particles.
This includes, in particular, all fermions.
The effective masses of such modes are $m\gsim \pi T$ so that there 
are no IR-problems but, according to Eq.(\ref{m0l:epsilon}), 
a small expansion parameter 
$\sim {g^2}/{\pi^2}$. The largest corrections are 
related to the top Yukawa coupling. 
After this step, the theory lives in 3d,
since all non-zero Matsubara modes have been removed. 
A second step can also be 
taken:~\cite{m0l:pert,m0l:generic,m0l:jp,m0l:mssmold} 
the zero components of the gauge fields
get radiatively a mass $m\sim gT$. Thus there are again 
no IR-problems but, according to Eq.~(\ref{m0l:epsilon}),
a small expansion parameter $\sim {g}/{\pi}$.

We will not discuss the precise details of these integrations
in any more detail here. For previous reviews, we refer 
to~\cite{m0l:erice}, and for explicit generic rules
for the integrations, to~\cite{m0l:generic}. 
For the logic behind the formal conjecture
concerning the parametric accuracy 
obtained with the types of super-renormalizable
effective theories discussed here 
(viz.,\ ${\delta G}/{G}\sim{\cal O}(g^3)$, where $G$ is 
a general non-vanishing bosonic Green's function), 
we refer to~\cite{m0l:parity}.

\section{Phase diagram for the Standard Model}

After the heavy degrees of freedom have been 
integrated out perturbatively, one still has to deal 
with the non-perturbative bosonic scales $m\lsim g^2T$. 
In the case of the Standard Model and many of its
extensions,~\cite{m0l:generic} the only 
infrared modes left are a single Higgs $\phi$ and 
the spatial SU(2) and U(1) 
gauge fields, with field strength tensors $F_{ij},B_{ij}$.
The Lagrangian is
\ba
{\cal L}_{\rm 3d} & = & 
{1\over2}\Tr F_{ij}^2 +{1\over4} B_{ij}^2+
(D_i\phi)^\dagger D_i\phi+m_3^2\phi^\dagger\phi+
\lambda_3(\phi^\dagger\phi)^2.
\label{m0l:lagr}
\ea
All knowledge about the original modes with
their physical zero temperature parameters, as well as about 
the temperature, 
is encoded in the expressions for the effective couplings 
$m_3^2,\lambda_3,g_3^2,g_3'^2$. 
Here $g_3,g_3'$ denote the SU(2) and U(1) gauge couplings. 

To get a feeling for the properties 
of the effective theory in Eq.~(\ref{m0l:lagr}), let 
us first apply 1-loop perturbation theory. Ignoring the 
tiny corrections from 
$g_3'^2/g_3^2$ for the moment 
and tuning $m_3^2$ to be on the phase transition line, we find 
a first order transition, with the discontinuity
\be
\Delta \frac{v}{gT} = \frac{1}{8\pi} 
\frac{g_3^2}{\lambda_3}\biggl[1 + {\cal O}(\frac{\lambda_3}{g_3^2})\biggr].
\label{m0l:disc}
\ee
(The ${\cal O}(\lambda_3/g_3^2)$ corrections 
are in fact not analytic.) 
Using that in the SM,
\be
\frac{\lambda_3}{g_3^2} \approx {1\over 8} \frac{m_H^2}{m_W^2} + 
{\cal O} \Bigl( \frac{g^2}{(4\pi)^2} \frac{m_\rmi{top}^4}{m_W^4}\Bigr),
\label{m0l:efflam}
\ee
and inspecting, say, $m_W(T) = gv/2$ in the broken phase, 
we however see from Eq.~(\ref{m0l:epsilon}) that for large (realistic) 
Higgs masses $m_H > m_W$ 
an IR sensitive computation of the type in Eq.~(\ref{m0l:disc})
cannot be trusted at all.

Fortunately, we do not need to rely on perturbation theory. The theory
in Eq.~(\ref{m0l:lagr}) is ideally suited for lattice simulations.
Due to the simplicity of the action and its low dimensionality, 
extrapolations to the infinite volume and continuum limits
can be systematically carried 
out~\cite{m0l:framework,m0l:contlatt,m0l:moore_a}.

%%%%%%%%%%%%%%%%%%%%%%%%%%%%%%%%% FIGURE %%%%%%%%%%%%%%%%%%%%%%%%%%%%%%%%%%
\begin{figure}[t!]

\centerline{\epsfxsize=5cm\hspace*{0cm}\epsfbox{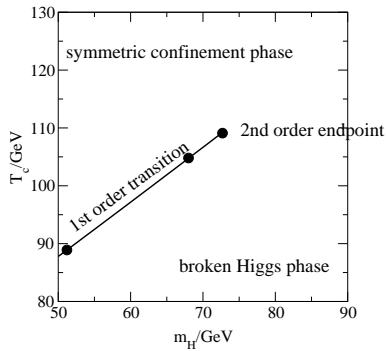}}

\caption[a]{The phase diagram of the physical Standard Model. The blobs
indicate simulation points (with errorbars), 
and the solid curve is a fit through them.}

\vspace*{10pt}

\label{m0l:fig:sm}
\end{figure}
%%%%%%%%%%%%%%%%%%%%%%%%%%%%%%%%%%%%%%%%%%%%%%%%%%%%%%%%%%%%%%%%%%%%%%%%%%%

The results of such lattice 
simulations from refs.~\cite{m0l:su2u1,m0l:ising,m0l:whatsnew} 
are summarised in Fig.~\ref{m0l:fig:sm}. There is a line of first order 
phase transitions, which however ends at a critical point, after which
there is only a crossover. The ending of the transition is in qualitative
contrast with the perturbative prediction in Eq.~(\ref{m0l:disc}). 

The  endpoint location has been determined with great precision, 
and corresponds in physical units to 
$m_{H,c}= 72.3(7)$ GeV, $T_c = 109.2(8)$ GeV~\cite{m0l:whatsnew}.
The errors here are statistical, remaining after continuum extrapolation.
In case one would now not be a priori convinced about the accuracy
of dimensional reduction at the percent level, we may note that 
a similar phase diagram exists also in the bosonic 4d
SU(2)$+$Higgs theory, where also direct 4d simulations 
have been carried out.~\cite{m0l:cfh}
While the statistical errors are slightly 
larger, the results agree completely with the results obtained  
after leaving out fermions from the dimensional reduction
expressions, as well as changing 
$g^2$ to the larger value used in ref.~\cite{m0l:cfh}.
A detailed comparison of the whole phase diagram has
been carried out in ref.~\cite{m0l:gr2}, with full agreement
for $m_H\gsim 30$ GeV.
Thus, we can really rely on the result in Fig.~\ref{m0l:fig:sm}. 

Let us now discuss the physical conclusions drawn from Fig.~\ref{m0l:fig:sm}.
As pointed out in the Introduction, 
for baryogenesis we need a first order phase transition, and 
even a strong one, $\Delta v/T\gsim 1$. We have learned that 
a first order transition only exists for $m_H < 72$ GeV. 
Furthermore, it has in fact $\Delta v/T \lsim  1.0$ down to
$m_H\sim 10$ GeV!~\cite{m0l:nonpert,m0l:ae,m0l:fh}
Thus, we observe that experimentally allowed Higgs masses 
$m_H \gsim 110$ GeV~\cite{m0l:lep} are very far from allowing
for a first order electroweak phase transition, let alone for 
electroweak baryogenesis. 

Finally, let us briefly mention a related physics topic, which 
is of some interest
because its solution is not yet understood
as well as the issues above.
It is the question as to how the properties of the phase 
transition change if there are primordial magnetic fields 
present,~\cite{m0l:bext} or, more generally, whether 
magnetic fields could in some way be related to the baryon 
asymmetry.~\cite{m0l:allor} 

\section{How to change the properties of the phase transition?}

{}From the discussion in the previous section, it is clear
that if we want to have an electroweak phase transition
which is of the first order for realistic Higgs masses, and even 
strongly so, then a drastic change is needed with respect
to the Standard Model. 

What is it that should be changed?
As the strength of the transition is determined by the scalar 
self-coupling, cf.\ Eq.~(\ref{m0l:disc}),
we apparently need some new degree of freedom which can decrease 
the effective $\lambda_3$ 
by ${\cal O}(100\%)$. To get such a large correction in the 
first place, we need a bosonic zero mode with an expansion 
parameter like in Eq.~(\ref{m0l:epsilon}). 
A simple perturbative 1-loop 
computation shows that at least the sign of loop 
corrections of this type is the correct one, 
negative. But to have an effect
of ${\cal O}(100\%)$, we need
$\epsilon \sim 1$, so that $m \sim g^2 T/\pi$.
That is, the degree of freedom should itself be non-perturbative!

In principle, there are two kinds of possibilities for such 
degrees of freedom. Either it is a new type of a gauge field, 
maybe something like technicolour,
or a scalar. In the latter, more natural
case, there are thermal corrections 
in the effective mass parameter, appearing as 
$m_3^2 \sim m_\rmi{4d}^2 + g^2T^2$. Thus, to get a total 
outcome of order $(g^2 T )^2$, a cancellation must take place 
such that the mass parameter $m_\rmi{4d}^2$ is negative. 
At zero temperature, the physical mass is then roughly
$m_\rmi{phys}^2 \lsim  m_\rmi{4d}^2 + m_\rmi{top}^2 \lsim m_\rmi{top}^2$. 
In order for such a relatively light degree of freedom not to have shown
up so far in precision electroweak observables, 
it should be an SU(2) singlet.

It is interesting that scalar degrees of freedom 
satisfying these requirements can be found in realistic theories. 
Consider, for instance, the MSSM. Potential candidates 
are squarks and sleptons. Since they should couple
relatively strongly to the Higgs, we should choose stops.

Now, the stops come left- and right-handed, $\tilde t_L, \tilde t_R$ 
(these states can also mix, but for lack of space we ignore this here).
The requirement of having a small violation of the 
electroweak precision observables means that the weakly 
interacting left-handed one cannot be light, $m_{\tilde t_L}\gg m_\rmi{top}$.
This helps also in getting a large Higgs mass
(see, e.g., ref.~\cite{m0l:erz}), 
\be
m_H^2 \sim  m_Z^2\cos^2\! 2\beta +  
\frac{3 g^2}{8 \pi^2} \frac{m_\rmi{top}^4}{m_W^2}
\ln\frac{m_{\tilde{t}_R}m_{\tilde{t}_L}}{m_\rmi{top}^2}.
\label{m0l:mH}
\ee
On the contrary, the SU(2) singlet stop $\tilde t_R$ can 
be ``light'', and serve as the desired new degree of 
freedom. (Then we of course lose half of the correction
to $m_H$ in Eq.~(\ref{m0l:mH}), but this is the price to pay.)
It should not be so light that the stop 
direction gets broken before the electroweak phase transition, 
though, because then one cannot get back to the SM minimum 
afterwards.~\cite{m0l:cms}

Another example of a viable light scalar degree of 
freedom is the complex gauge singlet of the NMSSM, 
which can also get broken during the transition
(for a recent study, see~\cite{m0l:hsch}). 

\section{Numerical results for the MSSM}

%%%%%%%%%%%%%%%%%%%%%%%%%%%%%%%%% FIGURE %%%%%%%%%%%%%%%%%%%%%%%%%%%%%%%%%%
\begin{figure}[t!]

\centerline{\epsfxsize=5cm\hspace*{0cm}\epsfbox{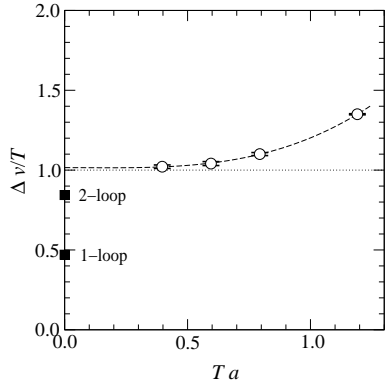}}

\caption[a]{The continuum extrapolation 
(i.e., lattice spacing $a\to 0$) of the numerically determined 
$\Delta v/T$, compared with perturbation theory,
for particular MSSM parameters.~\cite{m0l:cpsim} 
The transition is strongly of the first order here.}

\vspace*{10pt}

\label{m0l:fig:phasediag}
\end{figure}
%%%%%%%%%%%%%%%%%%%%%%%%%%%%%%%%%%%%%%%%%%%%%%%%%%%%%%%%%%%%%%%%%%%%%%%%%%%

Under the conditions described in the previous section, 
resummed 2-loop perturbation theory 
indicates that the electroweak phase
transition can indeed be strong enough 
for baryogenesis.~\cite{m0l:bjls}$^{-}$\cite{m0l:mlo3}
To check the reliability of perturbation theory, a 3d effective
theory was constructed in refs.~\cite{m0l:bjls,m0l:cpown}. 
The Lagrangian in Eq.~(\ref{m0l:lagr}) is supplemented by the 
right-handed stop field $U$ and the SU(3) gauge fields,  
with the field strength tensor $G_{ij}$:
\be
\delta {\cal L}_\rmi{3d} = {1\over 2} \Tr G_{ij}^2 + 
(D_i^s U)^\dagger D_i^s U + m_{U3}^2 U^\dagger U + 
\lambda_{U3}  (U^\dagger U)^2  + \gamma_3 \phi^\dagger \phi U^\dagger U. 
\label{m0l:delta}
\ee
Simulations with this action
have been carried out in ref.~\cite{m0l:mssmsim},
and with a more complicated version thereof, involving 
two Higgs doublets, in ref.~\cite{m0l:cpsim}.

The results are illustrated  
in~Fig.~\ref{m0l:fig:phasediag} for a particular choice
of parameters. We find the encouraging outcome that, in fact, 
for a light right-handed stop the 2-loop estimates 
are reliable and even somewhat conservative.
This is in strong contrast to the case of the 
SM at realistic Higgs masses. 

\begin{figure}[t!]

\centerline{\epsfxsize=5cm\hspace*{0cm}\epsfbox{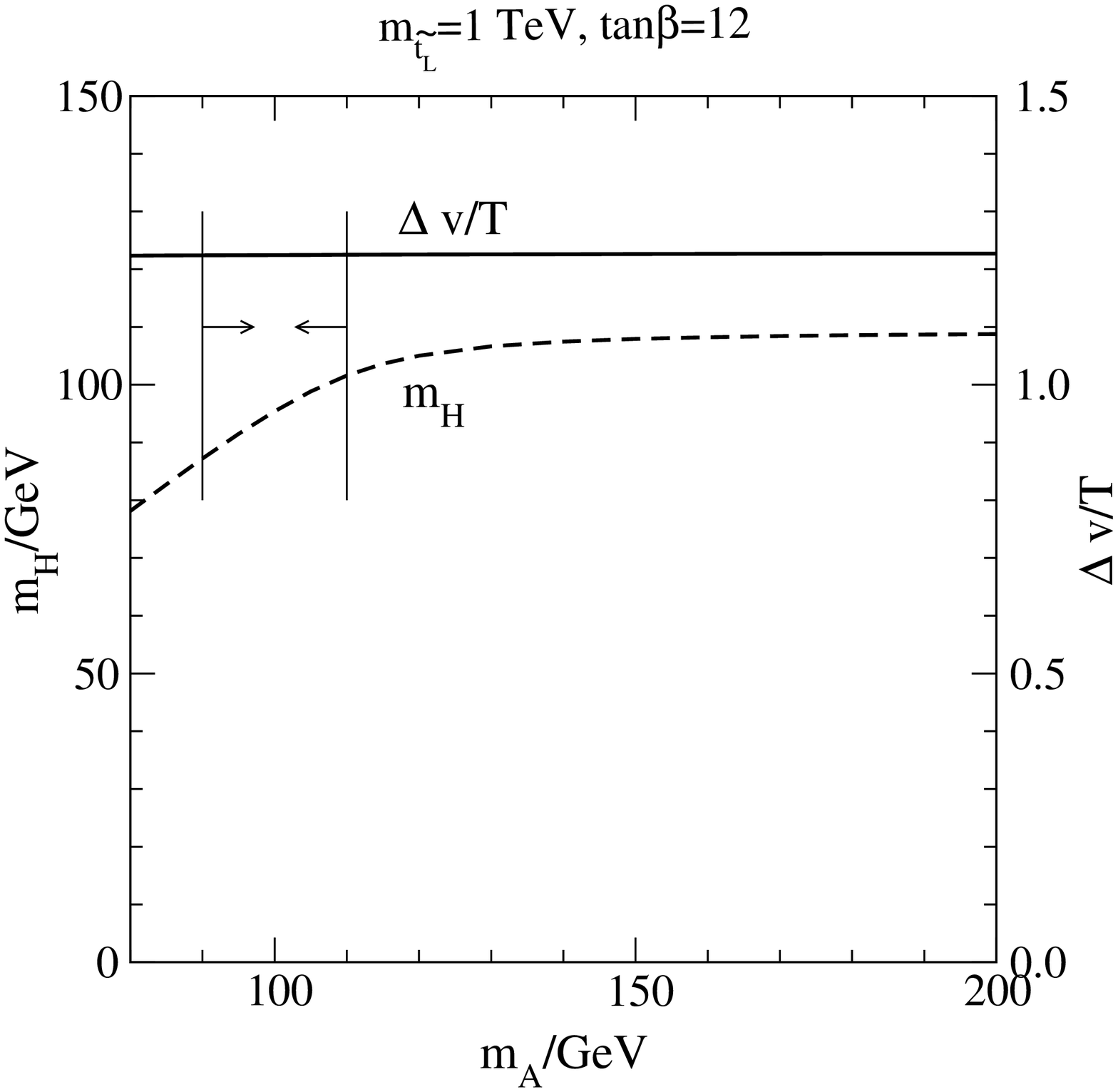}%
\epsfxsize=5cm\hspace*{1cm}\epsfbox{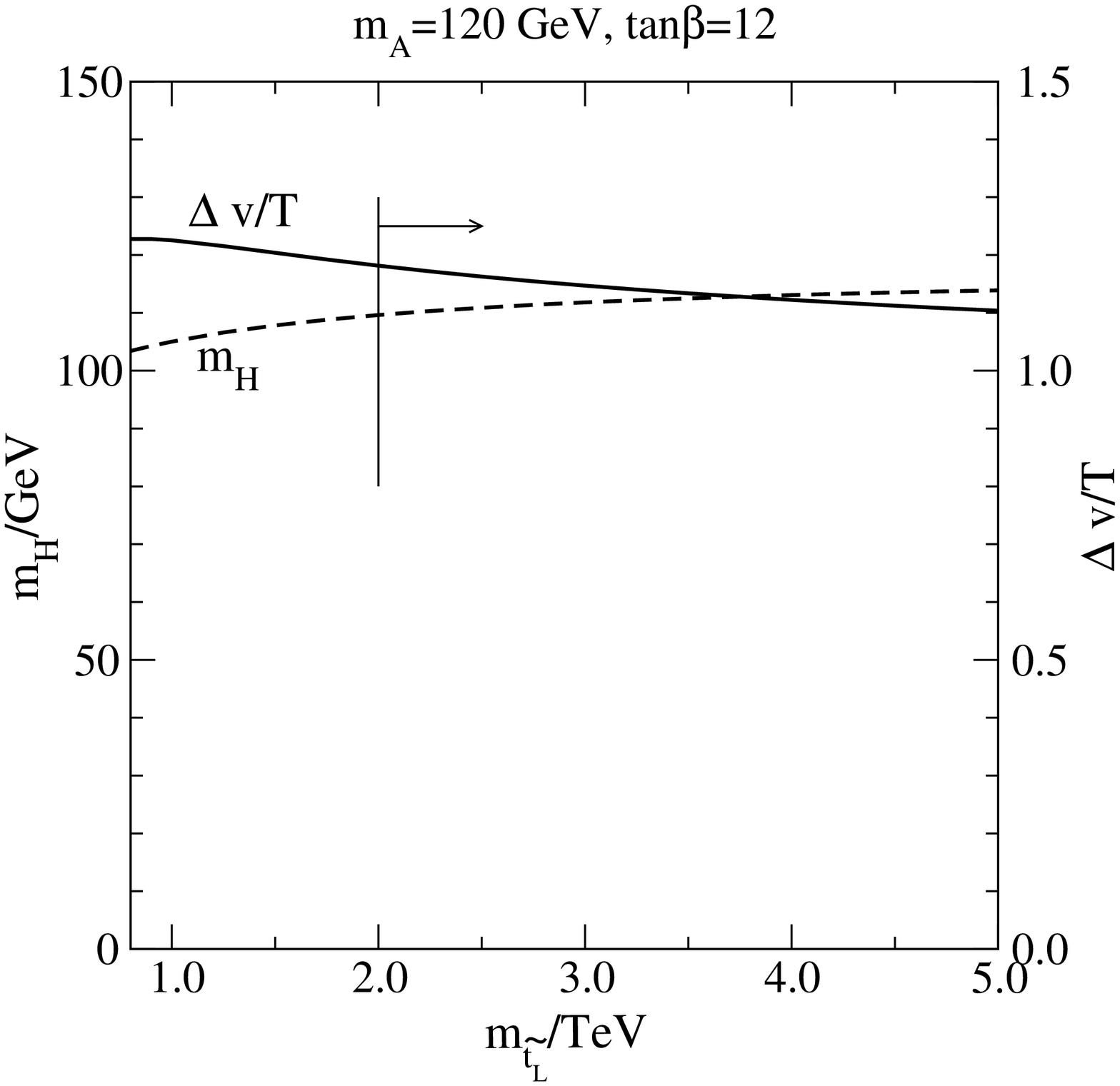}}

\caption[a]{The Higgs mass $m_H$, and the maximal discontinuity
$\Delta v/T$ according to 2-loop perturbation theory, as a function of 
$m_A$ (left) and $m_{\tilde t_L}$ (right).~\cite{m0l:cpsim} With lines
we indicate the regions least constrained by experiment
(already somewhat smaller than in ref.~\cite{m0l:lep}).}

\vspace*{10pt}

\label{m0l:fig:extensions}
\end{figure}

Next, we must ask more rigorously
whether this parameter region is indeed in agreement 
with all experimental data. The most important constraint 
comes from the lower bound on the Higgs mass.~\cite{m0l:lep} 
There is a parameter in the MSSM, $m_A$, 
which determines whether the Higgs sector resembles that in the 
Standard Model ($m_A \gsim 120$ GeV) or not ($m_A \lsim 120$ GeV).
In the latter case, the experimental lower bound is relaxed,~\cite{m0l:lep}
and since the transition needs not always get significantly weaker
(see Fig.~\ref{m0l:fig:extensions}, and ref.~\cite{m0l:cpsim} for more
details), this case is acceptable. The other possibility is having 
a large $m_A$, but then the left-handed stop should be quite heavy,
$m_{\tilde t_L}\gsim 2$ TeV (see Fig.~\ref{m0l:fig:extensions}),
in order to increase the Higgs mass.

%%%%%%%%%%%%%%%%%%%%%%%%%%%%%%%%% FIGURE %%%%%%%%%%%%%%%%%%%%%%%%%%%%%%%%%%
%%%%%%%%%%%%%%%%%%%%%%%%%%%%%%%%%%%%%%%%%%%%%%%%%%%%%%%%%%%%%%%%%%%%%%%%%%%

\section{CP violation in the background field configuration}

So far, we have discussed the strength of the phase transition.
Let us finally very briefly touch another issue relevant for 
baryogenesis, CP violation.

Even though the strength of the phase transition 
is determined by a single
``dynamical'' Higgs doublet, in the MSSM there are
of course in principle two of them, of which the one discussed above,
$\phi$ in Eqs.~(\ref{m0l:lagr}), (\ref{m0l:delta}), was
a particular linear combination (see, e.g., ref.~\cite{m0l:cpsim}).
However, the direction orthogonal to $\phi$ could play a role for 
CP violation. Indeed, when there are two Higgses, $H_1,\tilde H_2$, 
then there is for instance a new low-dimensional
gauge invariant CP-odd observable $\im H_1^\dagger \tilde H_2$.
It could obtain a non-vanishing value, in principle even 
a large one, as is the case in spontaneous CP violation.~\cite{m0l:lee} 
If so, then at the semi-classical level, fermion mass matrices 
may obtain non-trivial space-dependent phases, 
so that particles and anti-particles have different dispersion relations,
which could be of relevance to baryon asymmetry generation. 

However, in practice such
interesting effects are quite small. 
The reason is that all the effects related to the orthogonal 
direction are suppressed at least 
by $\sim g^2T/[\pi (m_A^2 + 0.5 T^2)^{{1\over 2}}]$.\cite{m0l:cpsim} 
Here $m_A$, experimentally $\gsim m_Z$,~\cite{m0l:lep} 
measures the deviation from the 
single Higgs doublet dynamics at zero temperature, 
and the temperature correction
is seen to increase the suppression further on. 

%%%%%%%%%%%%%%%%%%%%%%%%%%%%%%%%% FIGURE
\begin{figure}[t]

\centerline{\epsfxsize=5cm\hspace*{0cm}\epsfbox{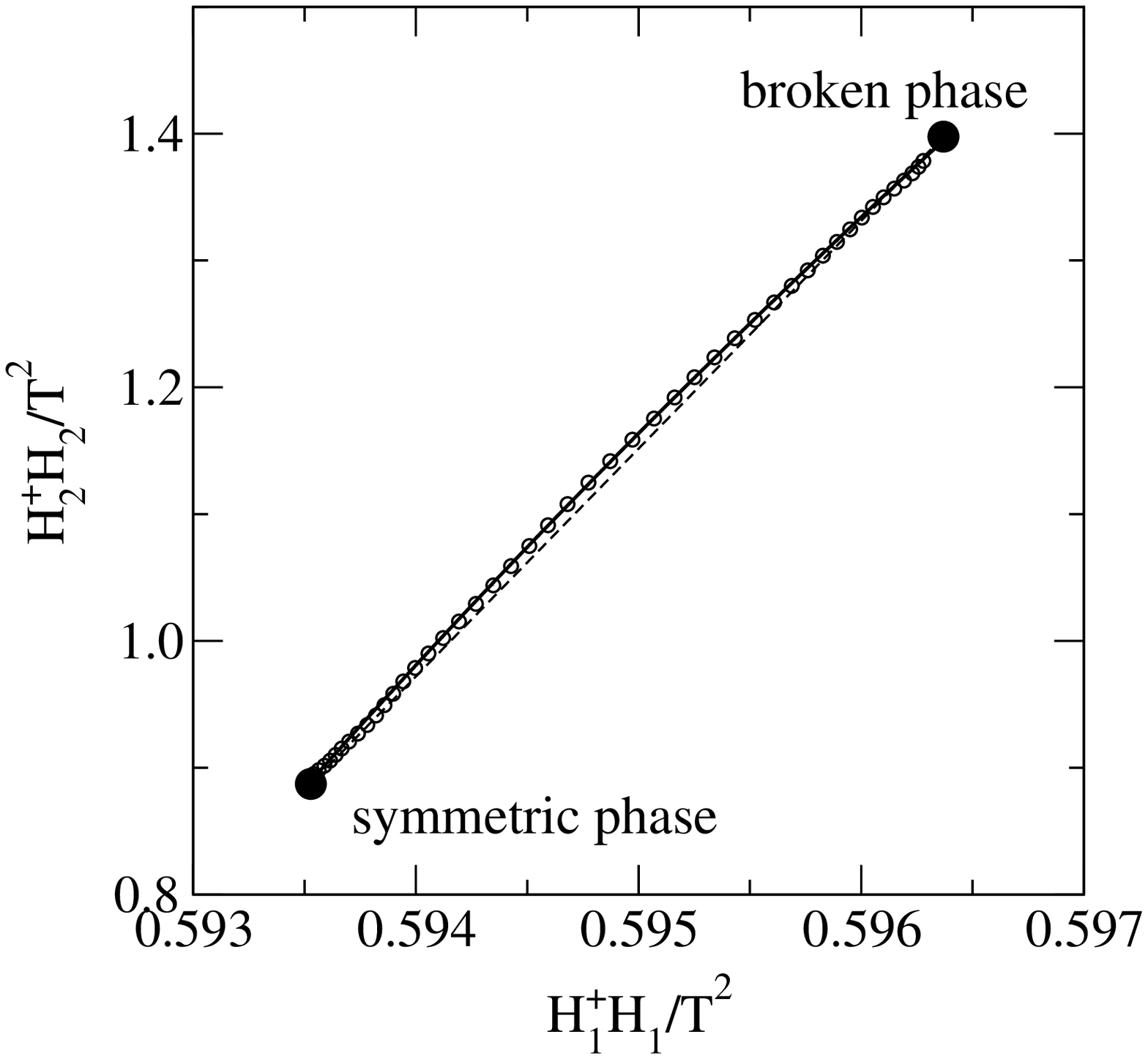}%
\epsfxsize=5cm\hspace*{1cm}\epsfbox{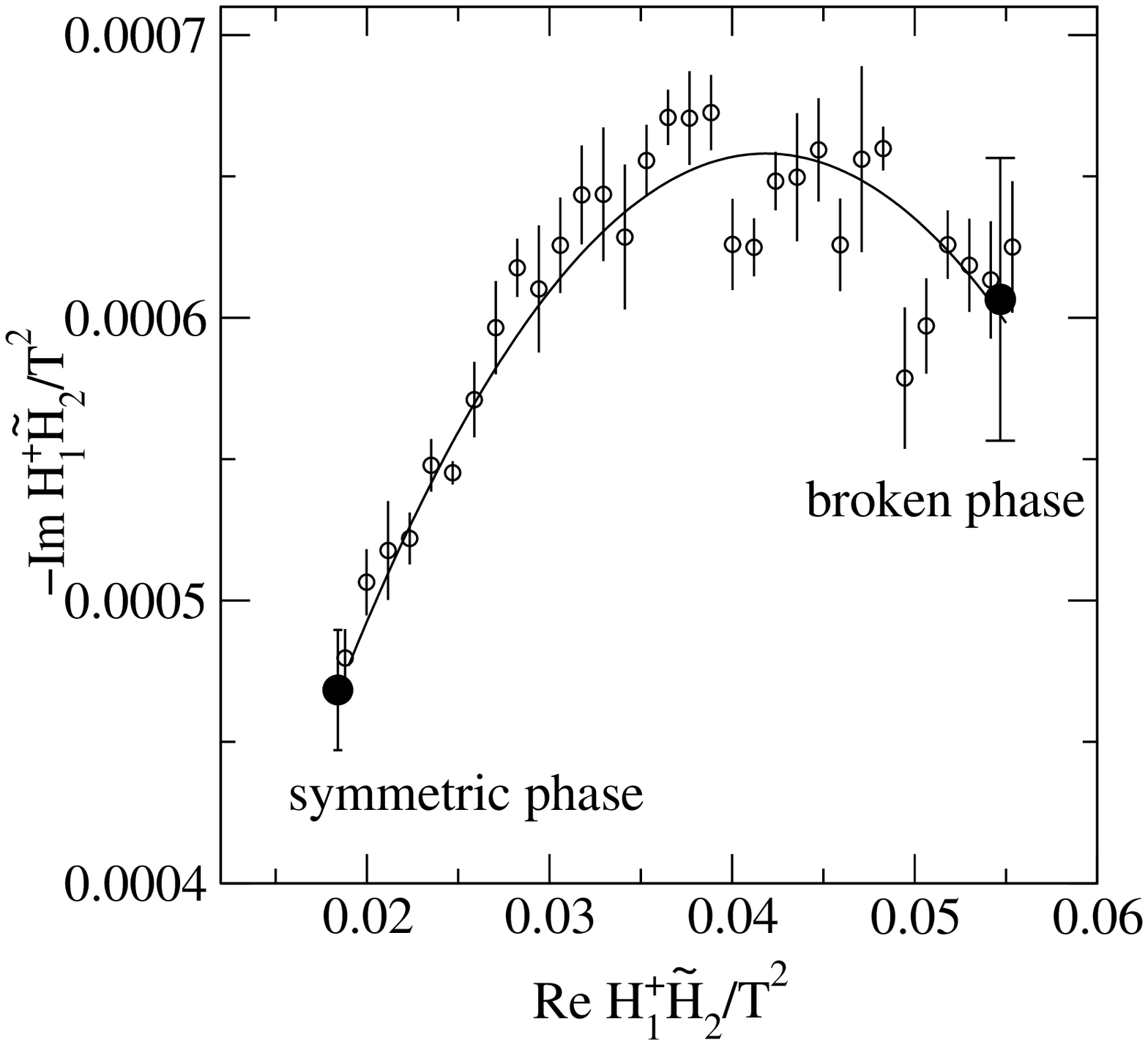}}

\caption[a]{Behaviour of 
$H_2^\dagger H_2$, plotted against $H_1^\dagger H_1$,
and $-\im H_1^\dagger \tilde H_2$, plotted  
against $\re H_1^\dagger \tilde H_2$, 
across a phase interface at $T_c$,
for particular MSSM parameters.~\cite{m0l:cpsim}} 
\label{fig:ReImb16}
\end{figure}
%%%%%%%%%%%%%%%%%%%%%%%%%%%%%%%%%%%%

This situation is illustrated in Fig.~\ref{fig:ReImb16}, which 
shows the profile of a stationary 
phase interface (for perturbative analyses of such objects, 
see ref.~\cite{m0l:pj} and references therein). 
We observe that in field space, 
the transition takes place in one particular direction
(see also~\cite{m0l:4dmssm}), and 
CP violation is small, as if the system effectively only had
one dynamical CP even Higgs doublet. 

\section{Conclusions}

We have reviewed 
the methods that can be employed in determining the 
thermodynamical properties of the phase transition related to electroweak
gauge symmetry restoration, as well as the 
numerical results that are
already available for the Standard Model  
and for its Minimal Supersymmetric extension. 
The general statement to be remembered,
is that there is no phase transition at all in  
electroweak theories with realistic Higgs masses, unless there is also 
another light scalar degree of freedom available, which plays
an essential role in the phase transition dynamics. 

{}As a concrete phenomenological example, 
we have recalled that it is still possible to find parts
of the MSSM parameter space, not excluded by experiment, where the electroweak
phase transition is strongly enough of the first order to allow for 
baryon asymmetry generation.  However, these parts are quite 
restricted: not only does one need one ``light'' stop, 
$m_{\tilde t_R} \lsim m_\rmi{top}$, in order to get a strong
transition, but also either another stop 
which is much heavier, $m_{\tilde t_L} \gsim 10\, m_\rmi{top}$, 
or a Higgs sector which has at least two independent light
particles, $m_A \lsim 120$ GeV, in order not to violate
experimental constraints.

\section*{Acknowledgments}

I thank D.~B\"odeker and K.~Rummukainen for useful comments. 
This work was partly supported by the 
TMR network {\em Finite Tempera\-ture Phase Transitions in 
Particle Physics}, EU contract no.\ FMRX-CT97-0122, and by 
the RTN network {\em Supersymmetry and the Early Universe}, 
EU contract no.\ HPRN-CT-2000-00152.


\begin{thebibliography}{99}

\bibitem{m0l:early}
D.A.~Kirzhnits,
%``Weinberg Model In The Hot Universe,''
{\it JETP Lett.}\  {\bf 15}, 529 (1972);
%%CITATION = JTPLA,15,529;%%
%
D.A.~Kirzhnits, A.D. Linde,
%``Macroscopic Consequences Of The Weinberg Model,''
{\it Phys.\ Lett.}\  B {\bf 42}, 471 (1972);
%%CITATION = PHLTA,B42,471;%%
%
S.~Weinberg,
%``Gauge And Global Symmetries At High Temperature,''
{\it Phys.\ Rev.}\  D {\bf 9}, 3357 (1974);
%%CITATION = PHRVA,D9,3357;%%
%
L.~Dolan, R.~Jackiw,
%``Symmetry Behavior At Finite Temperature,''
{\it Phys.\ Rev.}\ D {\bf 9}, 3320 (1974).
%%CITATION = PHRVA,D9,3320;%%

\bibitem{m0l:krs}
V.A.~Kuzmin, V.A.~Rubakov, M.E.~Shaposhnikov,
%``On anomalous electroweak baryon number 
%non-conservation in the Early Universe,''
{\it Phys.\ Lett.}\ B {\bf 155}, 36 (1985);
%%CITATION = PHLTA,155B,36;%%
%
M.E.~Shaposhnikov,
%``Baryon asymmetry of the Universe in Standard Electroweak Theory,''
{\it Nucl.\ Phys.}\ B {\bf 287},  757 (1987).
%%CITATION = NUPHA,B287,757;%%

\bibitem{m0l:rs}
V.A.~Rubakov, M.E.~Shaposhnikov,
%``Electroweak baryon number non-conservation
% in the early universe and in  high-energy collisions,''
{\it Usp.\ Fiz.\ Nauk} {\bf 166}, 493 (1996)
[hep-ph/9603208].
%%CITATION = HEP-PH 9603208;%%

\bibitem{m0l:nonpert}
K. Kajantie et al, %M. Laine, K. Rummukainen and M. Shaposhnikov,
%``The Electroweak Phase Transition: A Non-Perturbative Analysis,''
{\it Nucl.\ Phys.}\ B {\bf 466},  189 (1996)
[hep-lat/9510020].
%%CITATION = HEP-LAT 9510020;%%

\bibitem{m0l:moore_broken}
G.D. Moore,
%``Measuring the broken phase sphaleron rate nonperturbatively,''
{\it Phys.\ Rev.}\ D {\bf 59},  014503 (1999)
[hep-ph/9805264].
%%CITATION = HEP-PH 9805264;%%

\bibitem{m0l:ae}
P.~Arnold, O.~Espinosa,
%``The Effective potential and first order phase transitions:
% Beyond leading-order,''
{\it Phys.\ Rev.}\ D  {\bf 47}, 3546 (1993)
[hep-ph/9212235]; 
{\em ibid.}\ D {\bf 50}, 6662 (1994) (E).
%%CITATION = HEP-PH 9212235;%%

\bibitem{m0l:fh}
Z.~Fodor, A.~Hebecker,
%``Finite temperature effective potential to
% order g**4, lambda**2 and the electroweak phase transition,''
{\it Nucl.\ Phys.}\  B {\bf 432}, 127 (1994)
[hep-ph/9403219].
%%CITATION = HEP-PH 9403219;%%

\bibitem{m0l:pert}
K.~Farakos et al, %K.~Kajantie, K.~Rummukainen, M.~Shaposhnikov,
%``3-D physics and the electroweak phase transition: Perturbation theory,''
{\it Nucl.\ Phys.}\ B {\bf 425}, 67  (1994)
[hep-ph/9404201].
%%CITATION = HEP-PH 9404201;%%

\bibitem{m0l:e}
J.R. Espinosa, 
%``Dominant Two-Loop Corrections to the
% MSSM Finite Temperature Effective Potential,''
{\it Nucl.\ Phys.}\ B {\bf 475}, 273 (1996) [hep-ph/9604320];
%%CITATION = HEP-PH 9604320;%%
%
B. de Carlos, J.R. Espinosa,
%``The baryogenesis window in the MSSM,''
{\it Nucl.\ Phys.}\ B {\bf 503}, 24 (1997)
[hep-ph/9703212].
%%CITATION = HEP-PH 9703212;%%

\bibitem{m0l:bjls}
D.~B\"odeker et al, %P. John, M. Laine, M.G. Schmidt, 
%``The 2-loop MSSM finite temperature
% effective potential with stop  condensation,''
{\it Nucl.\ Phys.}\  B {\bf 497}, 387 (1997)
[hep-ph/9612364].
%%CITATION = HEP-PH 9612364;%%

\bibitem{m0l:cqw}
M.~Carena et al, %M.~Quir\'os, C.E.M.~Wagner,
%``Electroweak baryogenesis and Higgs and stop
% searches at LEP and the  Tevatron,''
{\it Nucl.\ Phys.}\  B {\bf 524}, 3 (1998) 
[hep-ph/9710401].
%%CITATION = HEP-PH 9710401;%%

\bibitem{m0l:cm}
J.M.~Cline, G.D.~Moore,
%``Supersymmetric electroweak phase transition:
% Baryogenesis versus  experimental constraints,''
{\it Phys.\ Rev.\ Lett.}\  {\bf 81}, 3315 (1998) 
[hep-ph/\\9806354].
%%CITATION = HEP-PH 9806354;%%

\bibitem{m0l:mlo2}
M.~Losada,
%``The two-loop finite-temperature effective
% potential of the MSSM and  baryogenesis,''
{\it Nucl.\ Phys.}\  B {\bf 537}, 3 (1999) [hep-ph/9806519].
%%CITATION = HEP-PH 9806519;%%

\bibitem{m0l:cms}
J.M.~Cline et al, %G.D.~Moore, G.~Servant,
%``Was the electroweak phase transition preceded by a color broken phase?,''
{\it Phys.\ Rev.}\ D {\bf 60}, 105035 (1999) [hep-ph/9902220].
%%CITATION = HEP-PH 9902220;%%

\bibitem{m0l:mlo3}
M.~Losada,
%``Mixing effects in the finite-temperature
% effective potential of the  MSSM with a light stop,''
{\it Nucl.\ Phys.}\  B {\bf 569}, 125 (2000) 
[hep-ph/9905441];
%%CITATION = HEP-PH 9905441;%%
%
S.~Davidson, T.~Falk, M.~Losada,
%``Dark matter abundance and electroweak baryogenesis in the CMSSM,''
{\it Phys.\ Lett.}\  B {\bf 463}, 214 (1999) 
[hep-ph/9907365].
%%CITATION = HEP-PH 9907365;%%

\bibitem{m0l:gpy}
A.D.~Linde,
%``Infrared problem in thermodynamics of the Yang-Mills gas,''
{\it Phys.\ Lett.}\ B {\bf 96}, 289 (1980);
%%CITATION = PHLTA,B96,289;%%
D.J.~Gross, R.D.~Pisarski, L.G.~Yaffe,
%``QCD and instantons at finite temperature,''
{\it Rev.\ Mod.\ Phys.}\ {\bf 53}, 43 (1981).
%%CITATION = RMPHA,53,43;%%

\bibitem{m0l:cfh}
F.~Csikor et al, %Z.~Fodor, J.~Heitger,
%``Endpoint of the hot electroweak phase transition,''
{\it Phys.\ Rev.\ Lett.}\  {\bf 82}, 21 (1999)
[hep-ph/9809291].
%%CITATION = HEP-PH 9809291;%%

\bibitem{m0l:4dmssm}
F.~Csikor et al, %Z.~Fodor, P.~Heged\"us, 
%A.~Jakov\'ac, S.D.~Katz, A.~Pir\'oth,
%``Electroweak phase transition in the MSSM:
% 4-dimensional lattice  simulations,''
{\it Phys.\ Rev.\ Lett.}\  {\bf 85},  932 (2000) [hep-ph/0001087].
%%CITATION = HEP-PH 0001087;%%

\bibitem{m0l:erice}
M.E.~Shaposhnikov,
%``Finite temperature effective theories,''
in {\it Erice 1996, Effective theories and fundamental interactions}, 
pp.\ 360--383 [hep-ph/9610247];
%%CITATION = HEP-PH 9610247;%%
%
A.~Nieto,
%``Perturbative QCD at high temperature,''
{\it Int.\ J.\ Mod.\ Phys.}\ A {\bf 12}, 1431 (1997)
[hep-ph/9612291];
%%CITATION = HEP-PH 9612291;%%
%
M.~Laine,
%``3D effective theories for the standard model and extensions,''
in {\it Eger 1997, Strong and electroweak matter '97}, 
pp.\ 160--177 [hep-ph/9707415].
%%CITATION = HEP-PH 9707415;%%

\bibitem{m0l:generic}
K. Kajantie et al, %M.~Laine, K.~Rummukainen, M.~Shaposhnikov,
%``Generic rules for high temperature dimensional
% reduction and their application to the standard model,''
{\it Nucl.\ Phys.}\  B {\bf 458},  90 (1996)
[hep-ph/9508379].
%%CITATION = HEP-PH 9508379;%%

\bibitem{m0l:gr2}
M.~Laine,
%``The renormalized gauge coupling and
% non-perturbative tests of  dimensional reduction,''
{\it JHEP} {\bf 9906}, 020 (1999)
[hep-ph/9903513].
%%CITATION = HEP-PH 9903513;%%

\bibitem{m0l:su2u1}
K.~Kajantie et al, % M.~Laine, K.~Rummukainen, M.~Shaposhnikov,
%``A non-perturbative analysis of the finite T phase 
% transition in  SU(2) x U(1) electroweak theory,''
{\it Nucl.\ Phys.}\ B {\bf 493}, 413 (1997)
[hep-lat/9612006].
%%CITATION = HEP-LAT 9612006;%%

\bibitem{m0l:owe}
O.~Philipsen et al, %M.~Teper, H.~Wittig,
%``On the Mass Spectrum of the SU(2) Higgs Model in 2+1 Dimensions,''
{\it Nucl.\ Phys.}\ B {\bf 469}, 445 (1996) [hep-lat/9602006].
%%CITATION = HEP-LAT 9602006;%%

\bibitem{m0l:gis}
M.~G\"urtler et al, %E.M.~Ilgenfritz, A.~Schiller,
%``Where the electroweak phase transition ends,''
{\it Phys.\ Rev.}\ D {\bf 56}, 3888 (1997) [hep-lat/9704013].
%%CITATION = PHRVA,D56,3888;%%

\bibitem{m0l:jp}
A.~Jakov\'ac, A.~Patk\'os,
%``Finite temperature reduction of
% the SU(2) Higgs model with complete static background,''
{\it Phys.\ Lett.}\  B {\bf 334}, 391 (1994)
[hep-ph/9405424];
%%CITATION = HEP-PH 9405424;%% 
%
%``Partial path integration of quantum fields:
% Two-loop analysis of the  SU(2) gauge-Higgs model at finite temperature,''
{\it Nucl.\ Phys.}\  B {\bf 494}, 54 (1997) 
[hep-ph/9609364].
%%CITATION = HEP-PH 9609364;%%

\bibitem{m0l:mssmold}
J.M.~Cline, K.~Kainulainen,
%``Supersymmetric Electroweak Phase Transition: Beyond Perturbation Theory,''
{\it Nucl.\ Phys.}\ B {\bf 482},  73 (1996) 
[hep-ph/9605235];
%%CITATION = HEP-PH 9605235;%%
%
%``Supersymmetric electroweak phase transition:
% Dimensional reduction  versus effective potential,''
{\em ibid.}\  {\bf 510}, 88 (1998) [hep-ph/9705201];
%%CITATION = HEP-PH 9705201;%%
%
M.~Losada,
%``High temperature dimensional reduction
% of the MSSM and other  multi-scalar models,''
{\it Phys.\ Rev.}\ D  {\bf 56},  2893 (1997) [hep-ph/9605266];
%%CITATION = HEP-PH 9605266;%%
%
G.R.~Farrar, M.~Losada,
%``SUSY and the electroweak phase transition,''
{\it Phys.\ Lett.}\ B {\bf 406}, 60 (1997)
[hep-ph/9612346];
%%CITATION = HEP-PH 9612346;%%
%
M.~Laine,
%``Effective theories of MSSM at high temperature,''
{\it Nucl.\ Phys.}\ B {\bf 481}, 43 (1996)
[hep-ph/9605283]; {\em ibid.}\ {\bf 548}, 637 (1999) (E).
%%CITATION = HEP-PH 9605283;%%

\bibitem{m0l:ll}
M.~Laine, M.~Losada,
%``Two-loop dimensional reduction and
% effective potential without  temperature expansions,''
{\it Nucl.\ Phys.}\  B {\bf 582}, 277 (2000) 
[hep-ph/0003111].
%%CITATION = HEP-PH 0003111;%%

\bibitem{m0l:parity}
K.~Kajantie et al, %M.~Laine, K.~Rummukainen, M.~Shaposhnikov,
%``High temperature dimensional reduction and parity violation,''
{\it Phys.\ Lett.}\  B {\bf 423}, 137 (1998)
[hep-ph/9710538].
%%CITATION = HEP-PH 9710538;%%

\bibitem{m0l:framework}
K.~Farakos et al, %K.~Kajantie, K.~Rummukainen, M.~Shaposhnikov,
%``3-d physics and the electroweak phase transition:
% A Framework for lattice Monte Carlo analysis,''
{\it Nucl.\ Phys.}\ B {\bf 442}, 317 (1995) [hep-lat/9412091].
%%CITATION = HEP-LAT 9412091;%%

\bibitem{m0l:contlatt}
M.~Laine,
%``Exact relation of lattice and continuum parameters
% in three-dimensional SU(2) + Higgs theories,''
{\it Nucl.\ Phys.}\ B {\bf 451}, 484 (1995)
[hep-lat/9504001];
%%CITATION = HEP-LAT 9504001;%%
%
M. Laine, A. Rajantie,
%``Lattice-continuum relations for 3d SU(N)+Higgs theories,''
{\it Nucl.\ Phys.}\  B {\bf 513}, 471 (1998)
[hep-lat/9705003].
%%CITATION = HEP-LAT 9705003;%%

\bibitem{m0l:moore_a}
G.D.~Moore,
%``Curing O(a) errors in 3-D lattice SU(2) x U(1) Higgs theory,''
{\it Nucl.\ Phys.}\ B {\bf 493}, 439 (1997)
[hep-lat/9610013];
%%CITATION = HEP-LAT 9610013;%%
%
%``O(a) errors in 3-D SU(N) Higgs theories,''
{\em ibid.}\   {\bf 523}, 569 (1998) [hep-lat/9709053].
%%CITATION = HEP-LAT 9709053;%%

\bibitem{m0l:ising}
K.~Rummukainen et al, %M.~Tsypin, K.~Kajantie, M.~Laine, M.~Shaposhnikov,
%``The universality class of the electroweak theory,''
{\it Nucl.\ Phys.}\  B {\bf 532}, 283 (1998)
[hep-lat/9805013].
%%CITATION = HEP-LAT 9805013;%%

\bibitem{m0l:whatsnew}
M.~Laine, K.~Rummukainen,
%``What's new with the electroweak phase transition?,''
{\it Nucl.\ Phys.\ B (Proc.\ Suppl.)}\  {\bf 73}, 180 (1999)
[hep-lat/9809045].
%%CITATION = HEP-LAT 9809045;%%

\bibitem{m0l:lep}
ALEPH, DELPHI, L3, OPAL Collaborations, \newline 
%presentation at \newline 
%``Rencontres de Moriond'', Les Arcs, France, March 11--25, 2000 \newline
http://alephwww.cern.ch/ALPUB/oldconf/oldconf00/29/moriond.ps.
%%CITATION = NONE;%%

\bibitem{m0l:bext}
P.~Elmfors et al, %K.~Enqvist and K.~Kainulainen,
%``Strongly first order electroweak phase transition
% induced by primordial  hypermagnetic fields,''
{\it Phys.\ Lett.}\  B {\bf 440}, 269 (1998)
[hep-ph/9806403];
%%CITATION = HEP-PH 9806403;%%
%
K.~Kajantie et al, %M.~Laine, J.~Peisa, K.~Rummukainen, M.~Shaposhnikov,
%``The electroweak phase transition in a magnetic field,''
{\it Nucl.\ Phys.}\ B {\bf 544}, 357 (1999) [hep-lat/9809004];
%%CITATION = HEP-LAT 9809004;%%
%
M.~Laine,
%``Vortex phases in condensed matter and cosmology,''
hep-ph/0001292.
%%CITATION = HEP-PH 0001292;%%

\bibitem{m0l:allor}
M.~Joyce, M.~Shaposhnikov,
%``Primordial magnetic fields, right electrons, and the Abelian anomaly,''
{\it Phys.\ Rev.\ Lett.}\  {\bf 79}, 1193 (1997)
[astro-ph/9703005];
%%CITATION = ASTRO-PH 9703005;%%
%
M.~Giovannini, M.E.~Shaposhnikov,
%``Primordial hypermagnetic fields and triangle anomaly,''
{\it Phys.\ Rev.}\ D {\bf 57}, 2186 (1998)
[hep-ph/9710234];
%%CITATION = HEP-PH 9710234;%%
%
D.~Comelli et al, %D.~Grasso, M.~Pietroni and A.~Riotto,
%``The sphaleron in a magnetic field and electroweak baryogenesis,''
{\it Phys.\ Lett.}\ B {\bf 458}, 304 (1999)
[hep-ph/9903227];
%
M.~Laine, M.~Shaposhnikov,
%``An all-order discontinuity at the electroweak phase transition,''
{\it Phys.\ Lett.}\ B {\bf 463}, 280 (1999) 
[hep-th/9907194].
%%CITATION = HEP-TH 9907194;%%

\bibitem{m0l:erz}
J.~Ellis, G.~Ridolfi, F.~Zwirner,
%``On radiative corrections to supersymmetric
% Higgs boson masses and their implications for LEP searches,''
{\it Phys.\ Lett.}\ B {\bf 262}, 477 (1991).
%%CITATION = PHLTA,B262,477;%%

\bibitem{m0l:hsch}
S.J.~Huber, M.G.~Schmidt,
%``SUSY variants of the electroweak phase transition,''
{\it Eur.\ Phys.\ J.}\  C {\bf 10}, 473 (1999) [hep-ph/9809506];
%%CITATION = HEP-PH 9809506;%%
%
%``Electroweak baryogenesis: Concrete in a SUSY model with a gauge  singlet,''
hep-ph/0003122.
%%CITATION = HEP-PH 0003122;%%

\bibitem{m0l:cpsim}
M. Laine, K.~Rummukainen,
%``Two Higgs doublet dynamics at the electroweak phase transition:
% A  non-perturbative study,''
hep-lat/0009025.
%%CITATION = HEP-LAT 0009025;%%

\bibitem{m0l:cpown}
M. Laine, K.~Rummukainen,
%``Higgs sector CP-violation at the electroweak phase transition,''
{\it Nucl.\ Phys.}\  B {\bf 545}, 141 (1999)
[hep-ph/9811369].
%%CITATION = HEP-PH 9811369;%%

\bibitem{m0l:mssmsim}
M.~Laine, K.~Rummukainen,
%``A strong electroweak phase transition up to m(H) approx. 105-GeV,''
{\it Phys.\ Rev.\ Lett.}\  {\bf 80}, 5259 (1998)
[hep-ph/9804255];
%%CITATION = HEP-PH 9804255;%%
%
%``The MSSM electroweak phase transition on the lattice,''
{\it Nucl.\ Phys.}\  B {\bf 535}, 423 (1998)
[hep-lat/9804019].
%%CITATION = HEP-LAT 9804019;%%

\bibitem{m0l:lee}
T.D. Lee, 
%``A Theory Of Spontaneous T Violation,''
{\it Phys.\ Rev.}\ D {\bf 8}, 1226 (1973).
%%CITATION = PHRVA,D8,1226;%%

\bibitem{m0l:pj}
S.J.~Huber et al, %P.~John, M.~Laine, M.G.~Schmidt,
%``CP violating bubble wall profiles,''
{\it Phys.\ Lett.}\  B {\bf 475}, 104 (2000) [hep-ph/9912278].
%%CITATION = HEP-PH 9912278;%%



\end{thebibliography}
\end{document}